\documentclass[]{article}
\usepackage{emulateapj}
\usepackage{graphicx}
\usepackage[usenames,dvips]{color}

\advance \voffset by -0.5cm\relax

\def\msun{{\rm ~M}_{\odot}}

\def\mpy{{\rm ~M}_{\odot} {\rm ~yr}^{-1}}

\begin{document}

\title{On The Maximum Mass of Stellar Black Holes}

\author{Krzysztof Belczynski\altaffilmark{1,2,3}, 
        Tomasz Bulik \altaffilmark{3,4},
        Chris L. Fryer\altaffilmark{1,5},
        Ashley Ruiter\altaffilmark{6,7},
        Francesca Valsecchi\altaffilmark{8},
        Jorick S. Vink\altaffilmark{9},
        Jarrod R. Hurley\altaffilmark{10}
}

\affil{
     $^{1}$ Los Alamos National Lab, 
            P.O. Box 1663, MS 466, Los Alamos, NM 87545 \\
     $^{2}$ Oppenheimer Fellow \\
     $^{3}$ Astronomical Observatory, University of Warsaw, Al. Ujazdowskie 4,
            00-478 Warsaw, Poland\\
     $^{4}$ Nicolaus Copernicus Astronomical Center, Bartycka 18, 00-716 
            Warsaw, Poland\\
     $^{5}$ Physics Department, University of Arizona, Tucson, AZ 85721\\
     $^{6}$ New Mexico State University, Dept of Astronomy,
            1320 Frenger Mall, Las Cruces, NM 88003\\
     $^{7}$ Harvard-Smithsonian Center for Astrophysics, 60 Garden
     St., Cambridge, MA 02138 (SAO Predoctoral Fellow)\\
     $^{8}$ Northwestern University, Dept of Physics \& Astronomy,
            2145 Sheridan Rd, Evanston, IL 60208\\
     $^{9}$ Armagh Observatory, College Hill, Armagh BT61, 9DG, Northern Ireland, United Kingdom\\
     $^{10}$ Centre for Astrophysics and Supercomputing, Swinburne University of Technology,
            Hawthorn, Victoria, 3122, Australia\\ 
     kbelczyn@nmsu.edu, tb@astrouw.edu.pl, clfreyer@lanl.gov,
     francesca@u.northwestern.edu,aruiter@nmsu.edu,jsv@arm.ac.uk,JHurley@groupwise.swin.edu.au
}

\begin{abstract}
We present the spectrum of compact object masses: neutron stars and 
black holes that originate from single stars in different environments. 
In particular, we calculate the dependence of maximum black hole mass on 
metallicity and on some specific wind mass loss rates (e.g., Hurley et 
al. and Vink et al.). Our calculations show that the highest mass black holes 
observed in the Galaxy $M_{\rm bh} \sim 15 \msun$ in the high metallicity
environment ($Z=Z_\odot=0.02$) can be explained with stellar models 
and the wind mass loss rates adopted here. To reach this result
we had to set Luminous Blue Variable mass loss rates at the 
level of $\sim 10^{-4} \mpy$ and to employ metallicity dependent 
Wolf-Rayet winds. With such winds, calibrated on Galactic black hole mass 
measurements,  the maximum black hole mass obtained for moderate 
metallicity ($Z=0.3\ Z_\odot=0.006$) is $M_{\rm bh,max} = 30 \msun$. 
This is a rather striking finding as the mass of the most massive known 
stellar black hole is $M_{\rm bh} = 23-34 \msun$ and, in fact, it 
is located in a small star forming galaxy with moderate metallicity. 
We find  that in the very low (globular cluster-like) 
metallicity environment the maximum black hole mass can be as high as 
$M_{\rm bh,max} = 80 \msun$ ($Z=0.01 Z_\odot=0.0002$). It is interesting to
note that X-ray luminosity from Eddington limited accretion onto an $80 \msun$ 
black hole is of the order of $\sim 10^{40}$ erg s$^{-1}$ and is comparable to
luminosities of some known ULXs. We emphasize 
that our results were obtained for single stars only and that binary 
interactions may alter these maximum black hole masses (e.g., accretion 
from a close companion). This is strictly a proof-of-principle study 
which demonstrates that stellar models can naturally explain even the 
most massive known stellar black holes.  
\end{abstract}


\section{Introduction}

Measuring the masses of celestial objects is one of the principal challenges 
in astrophysics. The primary method used to accomplish this task involves 
considering orbital motion. This standard method has been applied to a
number of X-ray binaries with the result showing that the mass function 
exceeds the maximal mass of a neutron star (NS), i.e.,  $2-3 \msun$, which points 
to the fact that there are systems containing black holes (BH). There are
also indirect ways of measuring black hole masses. The measurement of X-ray 
luminosity leads to a lower limit on the mass of the accreting body arising 
from the Eddington limit. This method, which was applied in the case of 
ultra luminous X-ray sources (ULX) has hinted that these systems are BHs 
with masses in excess of $100 \msun$ (Miller et al. 2004). This claim is 
subject to a number of assumptions, such as the isotropy of radiation (e.g.,
King 2009). Additionally, it was implied that BHs with mass smaller than 
$80 \msun$ can explain all, but the most luminous, ULXs (for most recent 
studies see Zampieri \& Roberts 2009; Kajava \& Poutanen 2009; Gladstone, 
Roberts \& Done 2009; Mapelli et al. 2009).
Nevertheless, there is a growing sample of stellar mass black holes with 
masses confirmed by observations of orbital binary motion, for a discussion 
see Pakull \& Mirioni (2003). Moreover, there are some solitary BH 
candidates which have been detected through microlensing experiments, and 
there are good prospects for observing more of such objects. A review of 
the observations of galactic BH candidates  (Orosz 2003; Casares 2007; 
Ziolkowski 2008) shows that the masses of stellar BHs range from a few solar 
masses to about $14 \pm 4 \msun$ (Greiner, Cuby \& McCaughrean 2001), in the 
case of GRS 1915-105, and $16 \pm 5$ for Cyg X-1 (Gies \& Bolton 1986).
However, in case of Cyg X-1 the estimates by different authors vary from 
$10 \msun$ (Herrero et al. 1995) to $20 \pm 5 \msun$ (Ziolkowski 2005).

The development of experimental techniques in recent years has allowed for 
the investigation of X-ray binaries in neighboring galaxies. This has led to 
the discovery of the largest stellar mass black hole in the binary IC10 X-1. 
The mass of the BH was found in the range of $23-34 \msun$ (Prestwich et al. 
2007; Silverman \& Filippenko 2008). It is interesting to note that the star 
formation rate in dwarf galaxy IC10 is very high, and its metallicity is low 
$\sim 0.3 Z_\odot$ (Massey et al. 2007). 

The spectrum of black hole masses is extremely interesting from the point 
of view of gravitational wave astronomy. The interferometric observatories 
like LIGO (Abramovici et al. 1992) and VIRGO (Bradaschia et al. 1990) can 
detect coalescences of $10 \msun$ BH binaries at a distance of $\approx 150$Mpc, 
and the sensitivity of the detectors will soon increase. Thus, the theoretical 
understanding of the distribution of masses of black holes formed by stars 
may soon be measurable with gravitational wave observations, and vice versa; 
knowledge of the BH mass spectrum may help to identify the parameter space 
which favors a high probability of source detection.  

Formation of BHs in the course of stellar evolution is connected with the end 
of the nuclear burning phase in a massive star. For the lower mass end of BH 
formation,  a meta-stable protoneutron star may be formed, and the black hole 
appears after accretion of the part of the stellar envelope that could not be 
expelled in the supernova explosion. In the case of the most massive stars the 
BH is formed through accretion of the entire stellar material (direct collapse or 
failed supernova). Thus the mass of the BH is determined mainly by the mass of 
the star at the moment the collapse takes place, as well as the explosion 
energy (e.g., Fryer 1999). The presupernova  mass is set predominantly by the 
amount of mass loss during stellar evolution. 

The mass loss for massive stars is due to stellar winds (single stars; e.g.,
Vink 2008) and additionally close interactions for stars that are found in 
multiple systems (e.g., binary stars; Hurley, Tout \& Pols 2002). 
Knowledge of the mass-loss rates for BH progenitors such as Wolf-Rayet (WR) 
stars -- massive stars found near the main-sequence (MS) losing mass at high 
rates and showing weak, or no, hydrogen lines in their spectra -- and 
Luminous Blue Variables (LBVs) -- extremely massive post-MS objects in a stage 
of evolution prior to becoming WR stars -- is therefore important for understanding 
BH masses. 
A subset of WR stars (those with no hydrogen lines) are the naked helium stars which 
can also be formed from less massive stars that lose their hydrogen envelopes 
on the giant branch or beyond. 

In this paper we review the recent results on modeling the stellar winds with 
a special emphasis on the metallicity dependence of these winds. This is a 
proof of principle study as we consider only the simplest case: that of single 
stellar evolution. We neglect the effects of binary interactions (e.g., mass 
loss/gain due to a close companion) and calculate the BH mass spectrum for 
single stars only. We combine the wind mass loss rates with the stellar 
evolution models and investigate the initial-remnant mass relation of stars for 
different metallicities. In particular, we calculate the maximum black hole 
mass for various models. The description of the wind mass loss rates and 
evolutionary model is presented in section \S\,2, \S\,3 contains the results 
of our calculations, and we summarize our findings in \S\,4.

\section{Model}

For our study we employ the single star evolutionary formulae of Hurley, 
Pols \& Tout (2000) that are used within the {\tt StarTrack} population synthesis 
code (Belczynski, Kalogera \& Bulik 2002; Belczynski et al. 2008). 
Updates to the Hurley et al. (2000) formulae in {\tt StarTrack} include 
an improved prescription for final remnant masses (see below) and the modeling 
of electron-capture supernovae where electrons are captured onto Mg atoms 
in an O-Ne-Mg stellar core leading to collapse to a NS (Podsiadlowski et al. 2004), 
thus extending the range of stellar masses that produce NSs (at the low-mass end). 
We also note that Hurley et al. (2000) fitted their stellar evolution formulae to 
detailed models of stellar masses of $100\, M_\odot$ or less and we extend this to 
include stars up to $150\, M_\odot$ -- the predicted maximum stellar mass for star 
formation under usual conditions (Weidner \& Kroupa 2004). However, the formulae 
are well-behaved within this extrapolation. Stars more massive than $150\, M_\odot$ 
which are now being considered as possible progenitors of intermediate-mass BHs in 
dense star clusters are not considered (see Glebbeek et al. 2009 for an overview). 

The employed single star evolutionary formulae were obtained for stellar models 
without mass loss. The formulae include the effects of mass gain (and potential 
rejuvenation) and mass loss on a star. Change of star mass may lead (e.g. for 
main sequence stars) to change of star central temperature and pressure, that 
affects the rate of nuclear reactions and a star lifetime. This leads also to 
the change of external star properties like its luminosity and radius. There 
is a feedback running from adopted wind mass loss rate (that depends on star 
properties; mass, luminosity and radius) to the star (mass loss affects star
mass and thus its radius and luminosity) and then back to the wind mass loss 
rate. This setup allows for employment of different wind mass loss rate 
prescriptions with the same set of underlying stellar models. The scheme that 
is used in our calculations is described in detail by Tout et al. (1997) and 
Hurley et al. (2002). This scheme is only an approximation in the mass
loss/gain treatment, and detailed stellar evolutionary calculations (e.g., 
Timmes, Woosley \& Weaver 1996; Limongi \& Chieffi 2006) with the specific set 
of winds could be, in principle, used to obtain the more accurate results. 
However, it is noted that both stellar models (e.g., mixing or reaction rates; 
Cassisi 2009) and wind mass loss rates (e.g., clumping or LBV phase; Vink 2008) 
are burdened with a number of uncertainties, rendering any estimate of a black 
hole mass a subject to large systematic errors.  

For stellar winds we use both the prescriptions given originally by 
Hurley et al. (2000: see \S\,2.1) and a new set of stellar winds (see \S\,2.2). 
In the descriptions of these winds  the following symbols are used: $L$  ($[L_\odot]$), 
$R$ ($[R_\odot]$), $M$ ($[M_\odot]$), and $T$ ($[K]$) for stellar luminosity, 
radius, mass and effective temperature, respectively, as well as $Z$ for 
metallicity with the solar value being $Z=Z_\odot=0.02$. 
The wind mass loss rates are denoted as $dM/dt$ ($[\mpy]$).

\subsection{Hurley et al. winds: previous reference model}

Here, as it is important for the presentation of the results 
and the discussion, we reiterate the 
wind mass loss prescriptions of the original source (Hurley et al. 2000) 
that have been used in {\tt StarTrack} over the last several 
years. The wind prescriptions were adopted as follows:
\begin{equation}
(dM/dt)_{\rm R} = 2 \times 10^{-13} {LR \over M} \mpy
\label{wind1}
\end{equation}
for stars on the Giant Branch and beyond (Kudritzki \& Reimers 1978; Iben \& 
Renzini 1983);
\begin{equation}
\log(dM/dt)_{\rm VW}= -11.4 + 0.0125 [P_0-100 {\rm max}(M-2.5,0)]
\label{wind2}
\end{equation}
for stars on the Asymptotic Giant Branch (Vassilidis \& Wood 1993),
with a maximum
value of $(dM/dt)_{\rm VW}= 1.36 \times 10^{-9} L \mpy$ and pulsation (Mira) 
period for these stars being $\log(P_0) =min(3.3,-2.07-0.9\log M+1.94\log
R)$;
\begin{equation}
(dM/dt)_{\rm NJ} = 9.6 \times 10^{-15} R^{0.81} L^{1.24} M^{0.16} 
                   \left({Z \over Z_\odot}\right)^{0.5} \mpy
\label{wind3}
\end{equation}
for luminous/massive ($L>4000L_\odot$) stars (Nieuwenhuijzen, H., \& de
Jager, C.\ 1990; Kudritzki et al. 1989);
\begin{equation}
(dM/dt)_{\rm WR1} = 10^{-13} L^{1.5}  \mpy    
\label{wind4}
\end{equation}
for WR stars (Hamann \& Koesterke 1998), it is noted that WR-like winds are
used also for stars with small H-rich mass envelope (WR star shines through), 
although in such a case the above formula is modified by the factor of 
$1-\mu$ where $\mu$ stands for (in first order approximation) fractional 
envelope mass (for details see equation 97; Hurley et al. 2000);
\begin{equation}
(dM/dt)_{\rm LBV1} = 0.1 (10^{-5} R L^{0.5}-1)^{3} 
                     \left({L \over 6\times 10^5}-1\right) \mpy            
\label{wind5}
\end{equation}     
for Luminous Blue Variables ($L>6 \times 10^5$ and $10^{-5} R L^{0.5} >1.0$);
Humphreys \& Davidson 1994). The LBV mass loss rate, if applicable, is added 
on top of stellar winds for H-rich stars, which in turn is calculated from 
$(dM/dt)={\rm max}[(dM/dt)_{\rm R}, (dM/dt)_{\rm VW}, (dM/dt)_{\rm NJ}]$.
For WR stars the wind is obtained from 
$(dM/dt)={\rm max}[(dM/dt)_{\rm R}, (dM/dt)_{\rm WR1}]$.

\subsection{Vink et al. winds: new reference model}

For hot massive H-rich stars (B/O spectral type) following Vink, 
de Koter \& Lamers (2001) we apply:
\begin{equation} 
\begin{array}{ll}
\log(dM/dt)_{\rm OB} = & -6.688+2.210 \log(L/10^{5}) \\
                       & -1.339 \log(M/30) -1.601 \log(V/2.0) \\  
                       & +0.85 \log(Z/Z_\odot) +1.07 \log(T/20000)\\
\end{array}
\label{wind10}
\end{equation}
with the ratio of wind velocity at infinity to escape velocity $V=v_\infty/v_{esc}=1.3$ 
for stars with $12 500 \leq T \leq 22500$K; and 
\begin{equation} 
\begin{array}{ll}
\log(dM/dt)_{\rm OB} = & -6.697+2.194 \log(L/10^{5}) \\
                       & -1.313 \log(M/30) -1.226 \log(V/2.0) \\ 
                       & +0.85 \log(Z/Z_\odot)+0.933 \log(T/40000) \\
                       & -10.92 [\log(T/40000)]^2 \\
\end{array}
\label{wind11}
\end{equation}
with $V=2.6$ for stars with $27 500 \leq T \leq 50 000$K.   
Around $T \sim 25 000$K there is a bi-stability jump that leads to rapid 
wind increase. The jump is due to recombination of the Fe IV to the Fe III ion which 
is a more effective line (wind) driver.  In the transition zone, we apply 
eq.~\ref{wind10} for $22500 \leq T \leq 25000$K and eq.~\ref{wind11} for 
$25000 \leq T \leq 27500$K.
We note that the above winds apply to stars that are more massive than 
$M_{\rm zams} \gtrsim 3-3.5 \msun$ (spectral type earlier than B8-7) as these
stars have $T>12500K$ at the Zero Age Main Sequence. 

For Luminous Blue Variable stars that are evolved (beyond the main sequence), H-rich
and extremely luminous ($L>6 \times 10^5$ and $10^{-5} R L^{0.5})>1.0$;
Humphreys \& Davidson 1994) we adopt  
\begin{equation}
(dM/dt)_{\rm LBV2} = f_{\rm lbv} \times 10^{-4} \mpy
\label{wind12}
\end{equation}
with the standard choice for a calibration factor $f_{\rm lbv}=1.5$ (see 
\S\,3.3 for justification). This formula is assumed to give the
full amount of LBV mass loss (unlike in the previous prescription in which 
LBV winds were added on top of the underlying massive star winds: \S\,2.1).
We note that this formula accounts both for LBV stellar wind mass loss as 
well as possible LBV shell ejections. Furthermore, it is assumed that LBV 
mass loss is independent of metallicity (see \S\,3.3 for further discussion).

For Wolf-Rayet (naked helium) stars we adopt  
\begin{equation}
(dM/dt)_{\rm WR2} = 10^{-13} L^{1.5} \left({Z \over Z_\odot}\right)^{m} \mpy
\label{wind13}
\end{equation}
which is a combination of the Hamann \& Koesterke (1998) wind rate estimate that 
takes into account WR wind clumping (reduced winds), and Vink \& de Koter 
(2005) wind $Z$-dependence who estimated $m=0.86$ for WR stars. 

For H-rich low mass stars, for which the above prescriptions do not apply, we
use Hurley et al. (2000) winds (\S\,2.1).

\subsection{Compact object masses: maximum BH mass}

To calculate the mass and type of the compact object remnant we use
Hurley et al.  (2000) single star evolutionary formulae with the
updated winds (see \S\,2.2) to obtain the presupernova star mass and
structure (i.e., star and CO core mass).  For a given CO core mass we
estimate the final FeNi core mass using evolutionary models of Timmes,
Woosley \& Weaver (1996). We use hydrodynamical calculations of
supernovae explosions to estimate the amount of fall back that may
occur during core collapse of the most massive stars (e.g., Fryer
1999; Fryer \& Kalogera 2001). Finally, to change baryonic mass to
gravitational mass of a remnant we use a prescription proposed by
Lattimer \& Yahil (1989) for neutron stars, and for black holes we
assume that the gravitational mass is $90\%$ of the baryonic mass.

Our prescriptions for the iron core mass, fallback mass and final
remnant mass are given in full detail in Belczynski et al. (2008, see their
\S\,2.3.1). For CO core masses below 5\,$M_\odot$, the remnant is set
to the iron core mass. For CO core masses above 7.6\,$M_\odot$, we
assume the star collapses directly to a black hole and its final mass
is equal to the star's total mass at collapse. In between these two
extremes, we use a linear fit between no fallback onto the iron core
and complete fallback. This fallback prescription is based on the 
explosion models from Fryer (1999) and is in agreement with
Fryer \& Kalogera (2001).  We can compare this prescription directly
to the latest analytic study of remnant masses by Fryer et al. (in preparation) 
using CO cores from the Heger et al. (2003) solar-metallicity progenitors.    
The maximum remnant masses for these solar metallicity progenitors between 
our prescription and this new analysis by Fryer et al. are within $\sim
10-20\%$ of each other. The differences between our prescription
and the Fryer et al. semi-analytic estimates are well within the
errors in current detailed stellar evolutionary models and collapse 
simulations.  
The final black hole mass is primarily set by the mass of the star at
collapse (our most massive remnants are produced in stars that collapse 
directly to a black hole). Differences in the pre-collapse star mass 
account for the bulk of the remnant mass differences between stellar 
models by different groups. These differences are primarily driven by 
different wind prescriptions. 

The most massive remnants are formed through direct collapse
where the entire (or nearly entire) presupernova star ends up
under the event horizon, contributing to the black hole mass. In those cases,
the explosion energy is too low to overcome the gravitational potential of an 
exploding star. For example, a Galactic ($Z=Z_\odot$) massive star with
an initial mass of $M_{\rm zams}=100 \msun$ is found at presupernova stage 
as a WR star with a mass of $M_{\rm sn}=11-15 \msun$ and a CO core mass of 
$M_{\rm co}=8-11 \msun$ and this star forms a black hole with 
mass of $M_{\rm bh}=10-13.5 \msun$. The lower masses in the above example
correspond to the Hurley et al. winds (\S\,2.1) and the higher masses to the new
adopted set of winds (\S\,2.2). In both cases the presupernova star was
massive enough to form a black hole through direct collapse. 
  
As we will show in the next sections, the change of stellar wind
prescriptions does not affect the compact object remnant masses for
stars with initial mass $M_{\rm zams} \leq 30 \msun$ (i.e., stars that
form neutron stars, and low mass black holes). For the highest mass
stars, the new wind prescription results in more massive presupernova
objects and heavier black holes.  Although the details of supernova
calculations are still rather uncertain, relevant studies (e.g., Fryer
1999) indicate that stars with very high initial masses $M_{\rm zams}
\geq 100 \msun$ form black holes through direct collapse or at least
with significant fall back (i.e., most of presupernova mass ends up in
the black hole).  Therefore, the maximum mass of a black hole, which
is the main subject of this study, will predominantly depend only on
{\em (i)} the employed stellar models and {\em (ii)} the adopted set
of stellar winds, both which set presupernova mass.

\section{Results}

\subsection{Standard Prediction}

\begin{figure}
\includegraphics[width=0.9\columnwidth]{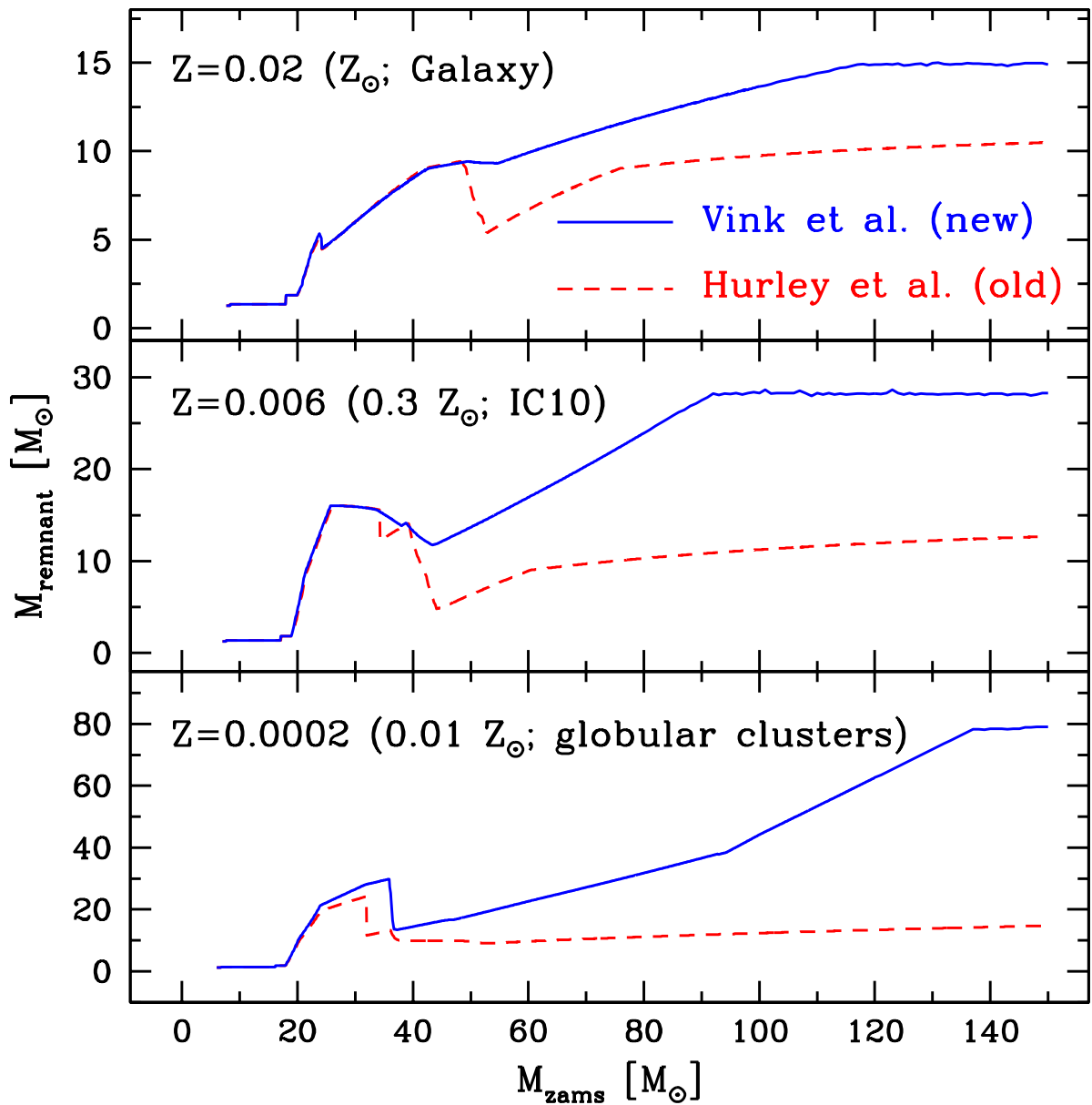}
\caption{
The initial-remnant mass relation for single stellar evolution for two wind 
prescriptions: the previously used Hurley et al. winds (\S\,2.1) and newly
adopted modified Vink et al. winds (imposed LBV winds at the level of $1.5 
\times 10^{-4} \mpy$; \S\,2.2).  
{\em Top panel:} Results for solar metallicity that correspond to
the stellar field populations in Galaxy. The predicted maximum black hole mass 
$M_{\rm bh,max} \sim 15 \msun$ for new and $\sim 10 \msun$ for old winds is 
consistent with the most massive stellar black holes observed in our Galaxy 
(e.g., in GRS 1915 $M_{\rm bh}=14 \pm 4 \msun$). 
{\em Middle panel:} Results for moderate metallicity that
correspond to stellar populations in galaxy IC10 which hosts the most
massive known stellar black hole ($M_{\rm bh}=23-34 \msun$). Note that
the predicted
maximum black hole mass $M_{\rm bh,max} \sim 30 \msun$ for new winds is 
consistent with the measurement in IC10, while $M_{\rm bh,max} \sim 15 
\msun$ obtained for old winds appears to be significantly too small. 
{\em Bottom panel:} Results for very low metallicity that
correspond to stellar populations of Galactic globular clusters or
metal-poor galaxies. The maximum black hole mass may reach $M_{\rm bh,max}
\sim 80 \msun$ or $\sim 25 \msun$ for the new and old wind prescriptions,
respectively. 
Note the change of vertical scale from panel to panel.
}
\label{Mfin1}
\end{figure}

Here we describe the predictions for the new adopted wind mass loss rates 
and we compare them with the previously employed prescription. 
The results for both new and old winds are presented in Figure~\ref{Mfin1} and
in Table~\ref{relation1}. 
In Figure~\ref{Mfin1} we present the initial-remnant mass relation for three
different metallicities; $Z=Z_\odot=0.02$, $Z=0.3\ Z_\odot=0.006$ and $Z=0.01\  
Z_\odot=0.0002$. The initial-remnant mass relation shows the compact object remnant 
mass ($M_{\rm remnant}$: either neutron star or a black hole) for a given 
initial ($M_{\rm zams}$: at Zero Age Main Sequence) star mass.
 
For high (Galaxy-like; $Z=Z_\odot=0.02$) metallicity, neutron star formation 
begins at $M_{\rm zams} =7.7 \msun$ with low mass NSs ($M_{\rm ns} =1.26 \msun$) 
formed through electron capture supernovae (e.g., Podsiadlowski et al. 2004),
while for higher initial masses (over $M_{\rm zams} = 8.3 \msun$; see 
Tab.~\ref{relation1}) NSs form through regular core collapse. 
In a rather wide range $8 \lesssim M_{\rm zams} \lesssim 18 \msun$ NSs are 
formed with $M_{\rm ns} =1.36 \msun$, then for $18 \lesssim M_{\rm zams}
\lesssim 20 \msun$ NSs are formed with $M_{\rm ns} =1.86 \msun$. There is
a bimodal final FeNi core mass distribution due to the mode of CO burning -- 
convective or radiative -- prior to core-collapse; Timmes et al. (1996).  
The latest results from the {\tt KEPLER} code do not show these same bimodal 
effect on the FeNi core mass, so this effect may not be real (Zhang et al. 2008). 
The detailed study of an alternative neutron star formation mass is underway 
(Fryer \& Belczynski, in preparation), we just note that the details of neutron 
star formation do not play crucial role in conclusions derived in this study. 
For $M_{\rm zams} \gtrsim 20 \msun$, fall back is expected to occur and this rapidly 
increases the mass of the remnant. Depending on the adopted limit for maximum NS mass 
($M_{\rm ns,max}$), black hole formation starts at $M_{\rm zams} \sim 20 \msun$ 
($M_{\rm ns,max}=2.0 \msun$) or at $M_{\rm zams} \sim 21 \msun$     
($M_{\rm ns,max}=3.0 \msun$). Note that the NS/BH transition is almost 
insensitive to the value adopted for the maximum NS mass as the initial-remnant mass
relation rises very steeply for the relevant remnant masses ($M_{\rm remnant} 
\gtrsim 2 \msun$). The steepness of the relation is due to the increasing
contribution of fall back in the final mass of the remnant. 

For lower metallicities, the general features of the initial-remnant mass relation
for neutron stars are very similar to the ones described above, with a natural
downward shift of initial masses, since lower metallicity stars lose less
mass in winds. In particular, for very metal-poor environments (globular 
cluster-like; $Z=0.01\ Z_\odot=0.0002$) NS formation starts at $M_{\rm zams}
=6.0 \msun$ and the transition to BH formation is found at $M_{\rm zams}  
=18 \msun$. Additionally, we note that whether we use the old or new wind 
prescription, the initial-remnant mass relation is virtually unaffected for
neutron stars. We therefore note that the wind mass loss rates presented by 
Hurley et al.(2000) do not differ significantly from the O/B winds presented 
by Vink et al. (2001) for neutron star progenitors ($M_{\rm zams} \sim 8-20 
\msun$). We can derive such a conclusion since in the considered mass
range, LBV and WR-like winds do not operate.  

For black holes various features of the initial-remnant mass relation change 
significantly with metallicity, and additionally the relation is different for 
the two sets of winds used. 
For $Z=Z_\odot=0.02$ the relation rises steeply for $M_{\rm zams} \sim 20
- 24 \msun$, and at the high end of this range there is a small dip after
which the relation resumes its rise but at a slower rate. This dip (a WR dip) 
corresponds to the onset of helium star winds; the threshold mass for removing 
the entire H-rich envelope is found at $M_{\rm zams,WR} = 24.2 \msun$ and above this 
mass the stars are subject to strong WR winds. At $M_{\rm zams,LBV} \sim 50 \msun$ 
we observe the onset of very strong LBV mass loss, and as the rates differ between old 
and new winds, the final-initial mass relations look different above the LBV
formation threshold. For the old rates, stronger LBV winds first lead
to a decrease (a LBV 
dip) and then a rather slow increase of BH mass with the initial star mass. For 
the new winds, the LBV winds are not as strong, and instead of a LBV dip there
is only a flattening-out at $M_{\rm zams} \sim 50 \msun$ and then the BH mass 
slowly increases with initial mass. The difference in relative strength of
LBV winds leads to noticeable (although not very drastic) differences
in black hole mass for the two models. Stars with $M_{\rm zams} \sim 100-150 \msun$ form 
$M_{\rm bh,max} \sim 10 \msun$ BHs in old models (high LBV mass loss), while 
they form $M_{\rm bh,max} \sim 15 \msun$ BHs in new models (moderate LBV mass 
loss). Note that at solar metallicity WR winds are the same for both models. 

For $Z=0.03\ Z_\odot=0.006$, BH masses are larger than in the case of solar
metallicity, as here winds are lower for smaller $Z$ so there is a larger mass 
reservoir at the time of BH formation. There are also significant qualitative 
differences in the shape of the relation. An LBV dip occurs for low initial mass 
($M_{\rm zams,LBV} \sim 34 \msun$), and in particular for old winds it is found 
even before the onset of the WR transition/dip ($M_{\rm zams,WR} = 37.4 \msun$), 
while for new winds it is found almost at the exact same place as the WR dip 
($M_{\rm zams,WR} = 33.8 \msun$; see Table 1). The shift of the onset of 
WR-like winds corresponds to decreasing mass loss with decreasing $Z$; 
therefore a higher initial stellar mass is required to form a naked helium star. 
Since at this $Z$ stars do not lose so much mass, they are more luminous and 
therefore they can reach the LBV phase at lower initial masses, so the LBV 
formation threshold moves the opposite way: to lower masses. 
For high initial masses, the BH mass becomes significantly different for
the two employed models. In particular, for the old set of winds, stars with
$M_{\rm zams} \sim 25-35 \msun$ and with $M_{\rm zams} \gtrsim 100 \msun$ 
form the most massive BHs: $M_{\rm bh,max} \sim 15 \msun$. With the new winds, stars 
with $M_{\rm zams} \gtrsim 100 \msun$ form maximum mass BHs: $M_{\rm bh,max} 
\sim 30 \msun$. 
This increasing difference in $M_{\rm bh,max}$ is due mostly to the dependence 
of WR winds on metallicity. For old winds, WR winds are assumed to be 
independent of $Z$ and therefore the increase in maximum BH mass is rather 
moderate. With the new prescription, WR winds are lower for lower $Z$ and we note 
a rather significant increase in maximum BH mass. To give an example, a star
with an initial mass of $M_{\rm zams} = 120 \msun$ loses $\sim 7$ and 
$\sim 25 \msun$ in WR winds for the new and old prescriptions, respectively 
(more examples in \S\,3.2). 

For $Z=0.01\ Z_\odot=0.0002$, the new model shows a new (subtle) feature. 
After the WR dip (at $M_{\rm zams}=36.9 \msun$) stars are a subject to strong 
WR winds, and the BH mass increases rather slowly with initial mass. However, 
at $M_{\rm zams}=92.5 \msun$ there is a small, but noticeable, steepening 
of the initial-remnant mass relation. At (and above) this point stars are 
sufficiently massive that the O/B winds and LBV winds are not strong enough 
to remove the entire H-rich envelope; so stars never become naked helium stars 
and are not subjected to strong WR winds. This is the main reason behind 
the formation of the most massive black holes at such low metallicity. The 
maximum mass BHs are found with $M_{\rm bh,max} \sim 80 \msun$ for 
$M_{\rm zams} \gtrsim 130 \msun$). At higher metallicities, the 
$Z$-dependent O/B type winds in conjunction with LBV mass loss remove 
the entire H-rich envelope, and the most massive stars are always a subject to WR 
winds. 
For the old model, the maximum BH mass is found to peak at $M_{\rm bh,max} \sim 25 
\msun$ for $M_{\rm zams} \sim 25-30 \msun$. In this mass range, stars are
not subjected to strong WR or LBV mass loss, but only to weaker $Z$-dependent 
winds (high BH mass since the winds are low for low $Z$; eq.~\ref{wind3}).
However, for higher initial masses stars lose most of their mass due to the 
effective and $Z$-independent LBV and WR winds (low BH mass since winds are high 
no matter what $Z$ is used; eq.~\ref{wind4} and eq.~\ref{wind5}).

\subsection{Examples of Mass Loss}

Here we give some numbers to indicate how much mass is lost in specific
winds for both models for stars that are relevant for maximum black 
hole mass calculations. 

We start with a star of $M_{\rm zams}=140 \msun$ at $Z=Z_\odot$. 
For Vink et al. winds, this star is $M_{\rm tms}=63 \msun$ at the end of
the main sequence (so it lost $\Delta_{\rm ms}=77 \msun$ in stellar winds during 
its main sequence evolution). When the star becomes a WR object its mass is 
$M_{\rm wr}=29 \msun$ (it lost $\Delta_{\rm lbv}=34 \msun$ in LBV winds). 
At the time of collapse the star's mass is $M_{\rm sn}=16.6 \msun$ (it lost
$\Delta_{\rm wr}=12.4 \msun$ in a WR wind) and it forms a (direct
collapse) BH of a mass $M_{\rm bh}=15 \msun$. 
For Hurley et al. winds, this star is $M_{\rm tms}=67 \msun$ at the end of
its main sequence evolution ($\Delta_{\rm ms}=73 \msun$). When the star becomes 
a WR object its mass is $M_{\rm wr}=28 \msun$ ($\Delta_{\rm lbv}=39 \msun$). 
At the time of collapse the star has $M_{\rm sn}=11.5 \msun$ (it lost 
$\Delta_{\rm wr}=16.5 \msun$ in WR wind) and there is a direct collapse to
a BH with $M_{\rm bh}=10.4 \msun$.
As said before, for solar-like metallicity the results are rather similar for
both sets of winds, and it can be concluded that most of the mass is lost 
during main sequence evolution ($\sim 53\%$).  The next phase of evolution removes 
less mass ($\sim 25\%$) through LBV winds, while finally WR winds deplete
the star of the relatively smallest (but still very significant) amount of mass 
($\sim 12\%$).

For comparison, we use the same initial mass of $M_{\rm zams}=140 \msun$ but
we shift to models with the very low metallicity $Z=0.01 Z_\odot$. 
For Vink et al. winds, this star is $M_{\rm tms}=134 \msun$ at the end of
the main sequence ($\Delta_{\rm ms}=6 \msun$). The LBV phase sets in shortly 
after the star leaves the main sequence and the star explodes during this phase 
at $M_{\rm sn}=86 \msun$ ($\Delta_{\rm lbv}=48 \msun$). 
Note that the star did not become a WR object
(i.e., the H-rich envelope was too massive to be removed by prior winds).  
The (direct) collapse leads to the formation of a BH with $M_{\rm bh}=78 \msun$.
For Hurley et al. winds, this star is $M_{\rm tms}=131 \msun$ at the end of
the main sequence ($\Delta_{\rm ms}=9 \msun$). When the star becomes a WR
object its mass is $M_{\rm wr}=58 \msun$ ($\Delta_{\rm lbv}=73 \msun$).
Note that in this prescription both main sequence winds and LBV winds are 
stronger and the star loses its entire H-rich envelope. 
At the time of collapse, the star's mass is $M_{\rm sn}=16 \msun$ as it lost an 
additional $\Delta_{\rm wr}=42 \msun$ in a WR wind and there is a direct 
collapse to a BH with $M_{\rm bh}=14.4 \msun$. 
The evolution is qualitatively different for both models because stronger winds 
in the Hurley et al. prescription allow for the formation of WR stars independent 
of metallicity, while for the Vink et al. winds 
stars avoid this phase at low metallicity and therefore they retain
more mass and form more massive BHs.
It is noted that for such low $Z$ only $\sim 5\%$ of the star's mass is lost in
winds during the main sequence, while LBV mass loss is estimated at the level of 
$\sim 34\%$ for Vink et al. and $\sim 52\%$ for Hurley et al.  
Additionally there is a WR mass loss of $\sim 30\%$ but only for the Hurley et al.
prescription.

We present here one more model at $Z=0.01 Z_\odot$ for $M_{\rm zams}=30 \msun$ 
as at this initial mass there is a peak in black hole mass (see \S\,3.1) for
the Hurley et al. winds. This star is $M_{\rm tms}=29.8 \msun$ at the end of
the main sequence ($\Delta_{\rm ms}=0.2 \msun$). After the main
sequence the star never
enters LBV nor the WR stage and at the time of collapse it has $M_{\rm sn}=25.8 
\msun$ (it lost $\Delta_{\rm pms}=4 \msun$ in post main sequence evolution). 
After direct collapse a BH is formed with $M_{\rm bh}=23.2 \msun$.  
Only very little mass is lost and most of it is lost during post main sequence
evolution ($\sim 13 \%$). For Vink et al. winds for the same initial
mass, the evolution is very similar with almost no mass loss during
the main sequence and with only a small amount lost during later evolutionary 
stages ($\Delta_{\rm pms}=0.5 \msun$).

\subsection{Effect of LBV Mass Loss}

\begin{figure}
\includegraphics[width=0.9\columnwidth]{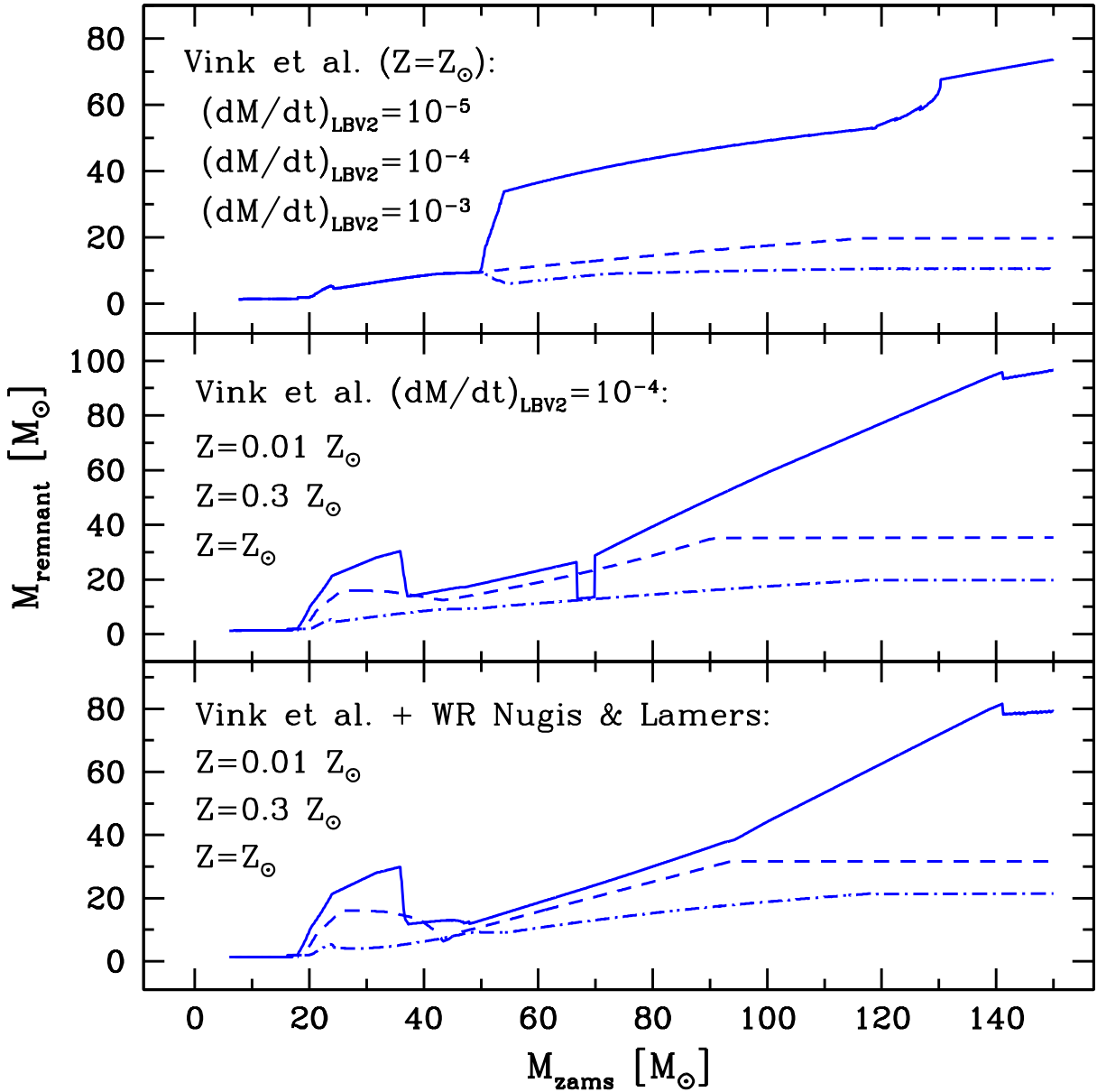}
\caption{
The initial-remnant mass relations for alternative parameter choices.  
{\em Top panel:} The relation for Vink et al. winds with the Galaxy-like 
metallicity ($Z=Z_\odot$) with three different assumed levels of LBV 
winds: $10^{-5} \mpy$ ($f_{\rm lbv}=0.1$, top line), $10^{-4} \mpy$ 
($f_{\rm lbv}=1.0$, middle line), $10^{-3} \mpy$ ($f_{\rm lbv}=10$, 
bottom line). The maximum BH mass is: $M_{\rm bh,max} \sim  80,\ 20,\ 10 
\msun$ for the top, middle and bottom lines, respectively. 
Note that Galactic black holes reach masses of $\sim 15 \msun$, implying 
the LBV winds at the level of $\sim 10^{-3}-10^{-4} \mpy$.
{\em Middle panel:} The relation for the Vink et al. winds but with lower
(than adopted in the reference model) LBV winds: $10^{-4} \mpy$.
Note the high maximum BH masses for this model $M_{\rm bh,max} \sim  100,\
35,\ 20 \msun$ for $Z=0.001,\ 0.3,\ 1.0  Z_\odot$, respectively.
{\em Bottom panel:} The relation for Vink et al. winds with modified WR
winds that were adopted from Nugis \& Lamers for these calculations.  
Note that results for low metallicity ($Z=0.01, 0.3 Z_\odot$, upper
curves) are almost
identical to our new reference model, while the high metallicity model 
results in a higher maximum BH mass ($M_{\rm bh,max} \sim 20 \msun$) than 
the reference calculation ($\sim 15 \msun$).} 
\label{Mfin2}
\end{figure}

In Figure~\ref{Mfin2} (the top panel) we show the initial-remnant mass relation for 
Vink et al. winds at solar metallicity ($Z=Z_\odot=0.02$) for three different 
levels of LBV mass loss. The top line shows the results for LBV winds of 
$(dM/dt)_{\rm LBV2} = 10^{-5} \mpy$ ($f_{\rm lbv}=0.1$). The maximum BH mass in 
this model is found to be $M_{\rm bh,max} \sim 80 \msun$, very high as the LBV 
winds are rather weak and the most massive stars never lose their entire H-rich 
envelope (i.e., WR winds do not turn on) and a significant fraction of the
star's initial mass is retained until the end of nuclear evolution.  
The middle line corresponds to a calculation with $(dM/dt)_{\rm LBV2} = 10^{-4} 
\mpy$ ($f_{\rm lbv}=1.0$) and the shape of the initial-remnant mass relation is 
very similar to the reference model that we have adopted (see Fig.~\ref{Mfin1}; top 
panel), but with a higher maximum BH mass of $M_{\rm bh,max} \sim 20 \msun$. 
Finally, the bottom line denotes a model with $(dM/dt)_{\rm LBV2} = 10^{-3} 
\mpy$ ($f_{\rm lbv}=10$) and this relation is almost identical to the one
that is calculated with our old wind prescription (Hurley et al.; \S\,2.1), 
although here we use a different set  of winds (Vink et al.; \S\,2.2). The 
maximum BH mass is $M_{\rm bh,max} \sim 10 \msun$.

We choose to set the strength of LBV winds in such a way that we can 
reproduce the most massive known BHs in the Galaxy: $\sim 15 \msun$ (e.g.,
Orosz 2003; Casares 2007; Ziolkowski 2008). It is found that for 
$f_{\rm lbv}=1.5$, which 
corresponds to a LBV mass loss at the level of $(dM/dt)_{\rm LBV2} = 1.5 
\times 10^{-4} \mpy$, the maximum BH mass reaches $15 \msun$ in our model 
with $Z=Z_\odot$ (which is appropriate for the Galaxy). 
As a consistency check, we test the same model (same LBV wind strength) for 
$Z=0.3\ Z_\odot$ which is the metallicity of the small star-forming galaxy IC10 
that hosts the most massive known BH in a binary:  X-1 $\sim 30 \msun$ 
(Prestwich et al. 2007; Silverman \& Filippenko 2008). As we can see in 
Figure~\ref{Mfin1} the model for IC10 can explain BH masses up to $\sim 30 \msun$.

The LBV winds were estimated at the level of $\sim 10^{-4} -- 10^{-5} \mpy$
by Vink \& de Koter (2002). Therefore, our choice of $(dM/dt)_{\rm LBV2} =
1.5 \times 10^{-4} \mpy$ is found very close to the high end of rate range 
derived by Vink \& de Koter (2002).
However, we demonstrated that for LBV mass loss rates as low as $\lesssim 10^{-5} 
\mpy$, black holes would reach unrealistically high masses $\gtrsim 80 \msun$
even for high (solar; $Z=Z_\odot=0.02$) metallicity (see Fig.~\ref{Mfin2};
top panel).  Although, such high mass BHs can not be excluded to exist in
the Galaxy, we choose to use the masses of the known BHs to calibrate 
our model. If more massive BHs are observed then it would be required to
lower our predicted LBV mass loss rates. 
We also speculate that the discrepancy may originate from the fact that the 
predicted rates do not take into account sporadic brightenings/eruptions of 
Luminous Blue Variables. If, during these episodes, some extra mass is lost
then obviously the predictions that account only for line/radiation driven
winds (as presented by Vink \& de Koter 2002) might be underestimated. 
For the extreme case of $\eta$ Carinae the amount of mass lost has been 
suggested to be substantial ($\sim 10 \msun$; Smith et al. 2003). However, 
it is as yet not clear how much mass (if any) is lost in LBV giant eruptions 
in general (e.g., van Genderen 2001).

We note that in our prescription for LBV mass loss, we assume the mass loss 
does not depend on metallicity. However, if most of the mass is lost via metal 
line-driving, we would expect there to be some dependence on metallicity.
It is plausible that such a metallicity dependence for the case of LBV winds 
is weaker than that of OB stars, as LBVs are closer to the Eddington limit, 
which plays an important role in the rate of mass loss (see Vink \& de Koter 
2002 for a more detailed discussion). It is clear that we need a better 
understanding of LBV mass loss before we can make any definitive conclusions 
with respect to the metallicity dependence of LBV mass loss. 

We also show the results for the Vink et al. winds but with slightly
lower LBV winds. In Figure~\ref{Mfin2} (the middle panel) we adopt
$(dM/dt)_{\rm LBV2}=10^{-4} \mpy$ and it is found that the maximum BH mass is: 
$M_{\rm bh,max} \sim  100,\ 35,\ 20 \msun$ for metallicities of $Z=0.001, 0.3, 
1.0\ Z_\odot$, respectively. This calculation naturally results in slightly 
higher BH masses, since the LBV winds are less effective in removing mass from 
stars. Note that this model can account for the potentially high mass estimate 
of the Galactic system Cyg X-1: $M_{\rm bh} = 20 \pm 5 \msun$, as well as 
the upper mass-range estimates for the black hole in IC10 X-1 $M_{\rm bh} 
\sim 34 \msun$ (e.g., Silverman \& Filippenko 2008).

\subsection{Wolf-Rayet winds}

Nugis \& Lamers (2000) have presented a study of WR wind mass loss rates and 
their dependence on WR star composition. Since, in the stellar models we 
employ (Hurley et al. 2000) the  actual composition of naked helium stars is
not available (there is only information on initial star metallicity), we can
only test the limited formula from Nugis \& Lamers (2000) that does not depend 
on the actual WR star composition
\begin{equation}
\log(dM/dt)_{\rm NL} = -5.73 + 0.88 \log M.
\label{wind20}
\end{equation}

In Figure~\ref{Mfin2} (the bottom panel) we show the initial-remnant mass
relation for our new adopted reference model (Vink et al.) in which we 
have replaced WR star winds (eq.~\ref{wind13}: Hamann \& Koesterke 1998; 
Vink \& de Koter 2005) with the rates given in eq~\ref{wind20} adopted 
from Nugis \& Lamers (2000). The results are presented for three 
different metallicities $Z=1.0, 0.3, 0.01 Z_\odot$. Although now WR 
winds do not depend on metallicity, there is still a metallicity 
dependence on H-rich winds (see eq.~\ref{wind10} and ~\ref{wind11}).  
The shape of these initial-remnant mass relations shows almost no differences 
with the new reference model results (see Fig.~\ref{Mfin1}). In fact, the
calculations using Nugis \& Lamers (2000) WR winds result in $M_{\rm bh,max} 
\sim 30, 80 \msun$ for low metallicities ($Z=0.3, 0.01 Z_\odot$,
respectively) -  
the same values as found in the new adopted model. For high metallicity 
($Z=Z_\odot$) the maximum BH mass is found to be $M_{\rm bh,max} \sim 20 \msun$ 
as compared to $\sim 15 \msun$ for the new model. If, in the model with 
Nugis \& Lamers (2000) WR winds, we required maximum mass to decrease to 
$\sim 15 \msun$ (maximum mass of known Galactic BHs) then we would have to 
adjust our calibration of LBV winds to $(dM/dt)_{\rm LBV2} = 7 \times 10^{-4} 
\mpy$.

\section{Discussion}

\subsection{Comments on Black Hole Masses}

A black hole in the Galactic system GRS 1915 ($M_{\rm bh}=14 \pm 4 \msun$) 
is found to exist in a binary with a low mass, evolved stellar companion 
($M_{\rm opt}=1.2 \pm 0.2 \msun$; K/M III; e.g., Greiner et al. 2001) 
which is filling its Roche lobe. If the companion star was born with such 
a low mass it could not possibly have significantly increased the mass of 
the black hole via mass transfer. Belczynski \& Bulik (2002) proposed an 
evolutionary scenario in which a companion star is born with a low mass 
and only after the black hole is formed it expands during its red giant 
evolution to fill its Roche lobe. In this scenario the observed/current 
mass of the black hole is virtually identical to the mass of the black 
hole at the time of its formation.  However, it was also argued that the 
initial mass of the companion may have been as high as $\sim 6 \msun$ and 
therefore, the black hole could have increased its mass by $\sim 4 \msun$ 
(Podsiadlowski, Rappaport \& Han 2003).

Another Galactic binary Cyg X-1 is host to a black hole ($M_{\rm bh}=16 
\pm 5 \msun$), and its massive companion ($M_{\rm opt} \sim 30 \msun$) is 
estimated to be close to, but not quite yet, filling its Roche lobe and is 
still on the main sequence (e.g., Gies \& Bolton 1986). The wind mass loss 
rate from the companion is measured at the level of $2.6 \times 10^{-6} 
\mpy$ (Gies et al. 2003). If a BH progenitor is a very massive star ($\sim 
100-150 \msun$) then its lifetime is $\sim 4$ Myr (e.g. Hurley et al. 
2000), and this sets the minimum age of the system. The main sequence 
lifetime of the companion is estimated to be $\sim 5$ Myr; we have assumed 
an initial mass of the companion $M_{\rm opt,zams} \sim 40 \msun$ as it 
has lost about $\sim 10 \msun$ in stellar winds over the course of the 
system's lifetime. This means that the black hole may have been accreting 
from its companion for only $\sim 1$ Myr. If we limit the accretion to the 
Eddington critical rate (which is $\sim 10^{-7} \mpy$ for a $15 \msun$ 
BH), the black hole in Cyg X-1 may have accreted only $0.1 \msun$. Had we 
relaxed the above assumption and allowed this BH to accrete at a much 
higher rate (e.g., Abramowicz et al. 1988; King 2002; Ohsuga 2007), the 
entire mass lost in stellar winds by the companion over $1$ Myr is $2.6 
\msun$. Only a fraction of this mass can be accreted by the black hole, 
even if the wind is focused in the orbital plane as suggested by some 
observational evidence (e.g., Herrero et al. 1995; Miller et al. 2005). 
This means that the black hole mass currently observed in Cyg X-1 is close 
to its formation mass. This must be true if the prior evolutionary history 
of Cyg X-1 does not include mass transfer via RLOF, and the only mass 
transfer proceeded through stellar winds from the companion (e.g., 
Ziolkowski 2005). Mass transfer via RLOF onto the BH in Cyg X-1 was 
considered by Podsiadlowski et al. (2003). It was concluded that if such a 
phase actually occurred, the black hole mass remained basically unchanged 
due to the fact that RLOF was proceeding on a very fast (thermal) 
timescale which is set by the massive companion, and most of the 
transferred mass was eventually lost from the system.

The most massive known extragalactic black hole ($M_{\rm bh} \sim 30 
\msun$) is in a binary system IC10 X-1 with a massive WR star companion: 
$M_{\rm opt} \sim 17-35 \msun$ with a mass at the high end of this range 
being the most likely case (Silverman \& Filippenko 2008). The WR star is 
well within its Roche lobe and the wind mass loss rate was estimated to be 
at the level of $10^{-5} - 4 \times 10^{-6} \mpy$ (Clark \& Crowther 
2004). Due to the short lifetime of such a massive WR star ($\sim 0.5$ 
Myr; e.g., Bulik, Belczynski \& Prestwich 2009), and since only a fraction 
of wind mass can be captured by the black hole, the increase in black hole 
mass due to wind accretion cannot exceed a few solar masses. Past 
evolution of this system has not yet been studied in detail and we cannot 
exclude the possibility that there was a phase with RLOF mass transfer 
from the unevolved companion (before it became a WR star). Such a mass 
transfer episode may have, in principle, significantly increased the black 
hole mass, and a thorough investigation of possible initial configurations 
of this system in conjunction with evolutionary calculations will be 
needed to clarify this issue (Valsecchi et al., in preparation).

In conclusion, it appears that the mass we have used to calibrate the 
maximum black hole mass for single star evolution in a Galactic-like 
environment (high metallicity; $Z=Z_\odot$): $\sim 15 \msun$ is consistent 
with the formation mass of Cyg X-1, which hosts one of the most massive 
BHs known in the Galaxy. With such a method, calibrated using the most 
recent metallicity-dependent wind mass loss rates, we predict that the 
maximum black hole mass (attained via single stellar evolution) is $M_{\rm 
bh,max} \sim 30 \msun$ for an environment of moderate metallicity, like 
that of the IC10 galaxy ($Z=0.3 Z_\odot$). Our prediction is rather 
striking as the mass of the most massive known BH -- that which resides in 
IC10 in binary X-1 -- is $M_{\rm bh}=23-34 \msun$. However, we cannot 
exclude the possibility that the BH in IC10 X-1 has accreted a significant 
amount of mass from its binary companion.

We also note that if our calibration of LBV winds is slightly modified 
(decreased) to $(dM/dt)_{\rm LBV2}=10^{-4} \mpy$, then we predict black 
holes (single stellar evolution) to form with masses up to $M_{\rm bh,max} 
\sim 20, 35, 100 \msun$ for $Z=Z_\odot$ (the Galaxy), $Z=0.3\ Z_\odot$ 
(IC10) and $Z=0.01\ Z_\odot$ (globular clusters), respectively (see 
Fig.~\ref{Mfin2}; middle panel). Therefore, our models can explain even 
the high end of the mass range estimates for the most massive known black 
holes.

\subsection{General Remarks}

In this study, we have analyzed two sets of stellar wind mass loss rates 
that are used in stellar evolution and population synthesis studies. In 
particular, we have adopted a revised and updated set of stellar winds for 
the population synthesis code {\tt StarTrack}. These updates will also be 
included as an option in the {\tt BSE} population synthesis code (Hurley 
et al. 2002). The previous stellar winds were adopted from the compilation 
of Hurley et al. (2000). The new adopted formulae employ metallicity 
dependent O/B star winds from Vink et al. (2001), reduced (clumping) 
Wolf-Rayet winds from Hamann \& Koesterke (1998) with metallicity 
dependence from Vink \& de Koter (2005) and Luminous Blue Variable mass 
loss rates that were calibrated in this study in such a way that stars 
could reproduce the most massive known black holes. In fact, our primary 
objective was to check whether within a given set of stellar evolutionary 
models and for available wind mass loss predictions, stars can form black 
holes that are as massive as those observed in our Galaxy ($\sim 14 \pm 4 
\msun$ for GRS 1915 and $\sim 16 \pm 5 \msun$ for Cyg X-1) and in external 
galaxies ($\sim 23-34 \msun$ for IC10 X-1). 

We have demonstrated that for solar metallicity environments (like our
Galaxy), both sets of winds - the previously employed (Hurley et al.) and 
the newly adopted (Vink et al.) - provide remnants massive enough to 
explain known Galactic black holes. However, the predictions are quite
different for intermediate and low metallicities. In summary, Hurley et 
al. winds remove too much mass from stars and within these models one 
does not expect the formation of black holes more massive than $\sim 25 
\msun$ even at very low metallicities. This is due to the combined effects of 
lack of metallicity dependence for WR winds and the rather strong LBV winds 
adopted in the Hurley et al. formulae. 
Quite contrary, for the Vink et al. winds, maximum black hole mass increases 
with decreasing metallicity, and for very low metallicity environments 
($Z=0.01 Z_\odot$) stars can from black holes as massive as $\sim 80 
\msun$. In particular, for a metallicity ($Z=0.3 Z_\odot$) of the galaxy
IC10 which hosts the most massive known stellar black hole, our newly 
adopted model predicts the maximum black hole mass of $M_{\rm bh,max}=30 
\msun$, which is consistent with the observed mass. 

Our new results bring the remnant mass predictions of {\tt StarTrack} more 
in line with the predictions of Fryer \& Kalogera (2001) and Heger et al. (2003).  
These studies argued for a strong metallicity dependence in agreement with the results by 
many groups studying massive stellar models:  Maeder \& Meynet (2001), 
Meynet \& Maeder (2003), Heger et al. (2002), Zhang et al. (2008).  

To summarize our results we show the dependence of maximum black hole mass
on metallicity in Figure~\ref{bhmax} for both sets of winds. 
The fact that the maximum black hole mass increases with metallicity
due to metal-dependent winds was recently investigated by stellar evolution
computations with metallicity-dependent main sequence and Wolf-Rayet mass loss by
Eldridge \& Vink (2006) where it was found that in lower metallicity galaxies
black holes are expected to have larger masses. Our results agree and to quantify
them we provide an approximate formula for the maximum black hole
\vspace*{0.2cm} mass:\\
$M_{\rm bh,max}=$\\
\begin{equation}
\left\{ \begin{array}{ll}
1403 \left({Z \over Z_\odot}\right)^2-548 {Z \over Z_\odot}+82.9 & {Z \over Z_\odot} < 0.11 \\
-23.7\left({Z \over Z_\odot}\right)^3+ 75.1 \left({Z \over Z_\odot}\right)^2-
84.4 {Z \over Z_\odot}+48 & {Z \over Z_\odot} \geq 0.11 \\
\end{array}
\right.
\label{bhmax}
\end{equation}
where the black hole mass is expressed in [$\msun$] and the relation holds true
for our newly adopted set of stellar winds (Vink et al.).  These formulae are only valid
for metallicities $0.01 \leq (Z/Z_\odot) \leq 1.5$; the range is set by 
the limits of the employed stellar evolutionary models (Hurley et al. 2000). The 
maximum black hole mass rises from $\sim 10 \msun$ for super-solar metallicity 
($Z=1.5 Z_\odot$) to $\sim 80 \msun$ for extremely low metallicity 
($Z=0.01 Z_\odot$). 
The Eddington limited accretion onto an $80 \msun$ black hole may give a rise to high 
X-ray luminosity $L_{\rm x} \sim 10^{40}$ erg s$^{-1}$ and $\sim 2 \times 10^{40}$ erg 
s$^{-1}$ in case of hydrogen and helium accretion, respectively. Such high
luminosities are comparable with X-ray emission of many known ULXs 
(e.g., Madhusudhan et al. 2008; Zampieri \& Roberts 2009).

For the most massive stars that will form maximum mass BHs, we have estimated 
the contribution of specific winds in the overall mass loss from a star.  
For high metallicities, most mass is lost during main sequence evolution
($\sim 50\%$), later LBV winds also remove a significant part of the star's 
initial mass ($\sim 25\%$) while finally the exposed naked helium star further 
loses mass ($\sim 10\%$) in WR-like winds.  
For very low metallicities, the majority of mass is lost during the LBV phase 
($\gtrsim 30\%$) with only small mass loss on the main sequence ($\sim 5\%$),
while the WR phase is not encountered (i.e., star never loses its H-rich
envelope). 
For intermediate metallicities, the mass is lost in comparable
quantities during the main sequence and LBV phases ($\sim 30\%$), with a small 
contribution to mass loss from WR-like winds ($\sim 5\%$).   

Our approach, based on one set of stellar evolution models by Hurley et al. (2000), 
matches observational constraints on black hole masses.  However, stellar
evolution models continue to produce very different results. The differences 
in evolution models produce uncertainties in stellar mass, core mass and
thus in final black hole mass. To illustrate these differences, we have compared 
our results based on the Hurley et al. (2000) formulae to models obtained 
with the EZ stellar evolutionary code (Paxton 2004) which we modified to include 
our specific wind mass loss prescriptions. The results of our comparisons are 
presented in Table~\ref{mod50} and ~\ref{mod100}. We note that the results, both 
for the analytic formulae and for detailed evolutionary calculations, show the 
same trends; for example the mass of the He core at the end of Main Sequence is 
higher for the old Hurley et al. winds as compared to models with the new Vink 
et al. winds.  But the EZ code typically predicts He core masses that are 
$\sim 20\%$ higher than those produced by our analytic formulae.   
Our analytic formulae were based on the revised Eggleton evolutionary 
code (Pols et al. 1998), while the EZ code is also based on the original Eggleton 
code but revised by Paxton (2004).  Even with very similar codes, properties 
in the core still have large differences. The shape of our black hole mass 
distribution is sensitive to these uncertainties.  However, the maximum 
black hole mass is produced in systems where the entire star collapses 
down to a black hole. To first order, the maximum black hole mass 
depends upon the final stellar mass and not on the internal stellar structure.  
The final stellar mass depends mostly on the wind mass loss rates, but it
changes also with the adopted stellar evolutionary model (which results in
different luminosities, radii, etc. so the same mass loss prescription gives
a different rate). Thus, our results are in the end limited by the details of
stellar evolutionary modeling uncertainties. It would be very useful if the
stellar evolutionary community could provide the final stellar masses that
are obtained with different codes. Such a comparison would set the
uncertainty of our result on the final black hole mass. One can hope that the 
final stellar masses obtained with various detailed evolutionary codes but
with the same set of wind mass loss rates would not differ by more than
$\sim 30\%$. 

Some stellar evolutionary models employ rotation (Heger et al. 2000; Maeder \& 
Meynet 2001; Meynet \& Maeder 2003; Vazques et al. 2007; de Mink et al. 2009). 
For stars that rotate fast (either low
metallicity single stars or tidally locked stars in close binaries), the effect 
of rotation may play an important role on their evolution and subsequent formation of 
the compact object remnant. The primary effect of rotation is mixing of 
additional material into the central part of the star where burning is taking
place, thus leading to the formation of a more massive core (as compared with 
non-rotating models). This should lead, in principle, to the formation of more 
massive remnants (e.g., it should increase the maximum black hole mass). 
However, rotation makes a star oblate, increasing the temperature of the polar 
regions. Thus, there is more radiation to drive winds, and the
net effect may be an increase in mass loss from a rapidly spinning star leading to  
a lower presupernova mass and so, reduction of the final compact object mass. It is 
not completely clear what would be the full effect of rotation on the maximum black hole 
mass, especially since the initial rotation of massive stars is not known, and 
this is particularly true for low metallicities.   

Finally, we want to emphasize that our conclusions are obtained only for 
single star models. In other words we do not take into account potential
effects of binary evolution. Those, in principle, may increase the mass of 
a black hole (via accretion from a close companion) or reduce its mass (by
mass loss from its progenitor). Additionally, the black hole masses that 
are known come from the observations of very specific types of sources:  
X-ray binaries; and thus it may be possible that either some evolutionary 
and/or observational selection effects hide the true, intrinsic maximum 
black hole mass. Therefore, by its nature, this is only a proof-of-principle 
study demonstrating that with some (widely used) evolutionary models and 
a set of (updated) metallicity-dependent stellar winds, regular single 
star evolution can naturally explain the metallicity-dependence of the most 
massive black holes that are known today. Additionally, it needs to be
highlighted that all three major components used in our study {\em (i)}
stellar evolution models, {\em (ii)} wind mass loss rate predictions and
{\em (iii)} core collapse/supernova calculations that are {\em all} very
important in remnant mass estimates are burdened with significant
uncertainties and sometimes even unknowns. We have attempted to collect
some of the most recent and widely used results (both observational and
theoretical) to provide a self-consistent physical model for remnant mass
calculations. However, we note that this model is still subject to future
adjustments and changes as more constraints will become available.

\acknowledgements
We express special thanks to Vicky Kalogera for critical comments. 
KB and TB acknowledge the support from KBN grant N N203 302835. 
This work was carried out in part under the auspices of the National Nuclear
Security Administration of the U.S. Department of Energy at Los Alamos
National Laboratory and supported by Contract No. DE-AC52-06NA25396.

\clearpage

\begin{deluxetable}{lccccccccr}
\tablewidth{420pt}
\tablecaption{Characteristic Properties of Initial-Remnant Mass Relations\tablenotemark{a}}
\tablehead{
      &   &       &              & $M_{\rm zams}$[$\msun$]: &              &    &     & $M_{\rm remnant}$: \\
      &   & WD/NS\tablenotemark{b} &      NS/BH   &      NS/BH               &        NS/BH & WR & LBV & $M_{\rm bh,max}$\\ 
Model & Z &       & ($2.0\msun$) & ($2.5\msun$)             & ($3.0\msun)$ &    &     & [$\msun$]  
} 

\startdata
Hurley & $Z_\odot$       & 7.7 (8.3) & 20.2 & 20.8 & 21.3 & 24.2 & 48.5 & 10.5 \\
Hurley & $0.3\ Z_\odot$  & 7.0 (7.7) & 19.2 & 19.4 & 19.6 & 37.4 & 34.3 & 16.0 \\
Hurley & $0.01\ Z_\odot$ & 6.0 (6.8) & 18.0 & 18.2 & 18.3 & 36.2 & 32.0 & 24.2 \\
       &                 &     &      &      &      &      &      &      \\
Vink   & $Z_\odot$       & 7.7 (8.3) & 20.2 & 20.8 & 21.2 & 24.2 & 49.7 & 15.0 \\
Vink   & $0.3\ Z_\odot$  & 7.0 (7.7) & 19.0 & 19.2 & 19.4 & 33.8 & 33.9 & 28.3 \\ 
Vink   & $0.01\ Z_\odot$ & 6.0 (6.8) & 17.9 & 18.1 & 18.2 & 36.9\tablenotemark{c} & 32.0 & 79.1 \\
\enddata
\label{relation1}
\tablenotetext{a}{Initial star mass ($M_{\rm zams}$) is given for:
transition of WD to NS formation and NS to BH formation for three assumed
maximum NS masses $M_{\rm ns,max}=2.0,\ 2.5,\ 3.0 \msun$. Minimum initial mass 
over which Wolf-Rayet star forms and over which Luminous Blue Variable forms. 
In the last column we list the maximum BH mass that is formed in a given model.} 
\tablenotetext{b}{Number in parenthesis: initial star mass for transition from 
electron capture supernova NS to core collapse NS.} 
\tablenotetext{c}{WR stars form only in the limited range of $M_{\rm zams}=36.9
-92.5 \msun$; for details see\S\.3.1.} 
\end{deluxetable}
\clearpage

\begin{deluxetable}{lcccc}
\tablewidth{420pt}
\tablecaption{Comparison of Models: $M_{\rm zams}=50 \msun$ \tablenotemark{a}}
\tablehead{
$Z$/Wind& $M_{\rm tms}$[$\msun$] & $M_{\rm he}$[$\msun$] & $\tau_{\rm ms}$[Myr] & $M_{\rm bh}$[$\msun$] }

\startdata
0.02/NoWind   & 50.0 (50.0) & 22.5 (18.9) & 3.8 (4.2) & (45.0) \\
0.02/Hurley   & 42.3 (41.3) & 21.2 (14.6) & 3.9 (4.3) & (7.9)  \\
0.02/Vink     & 42.3 (40.6) & 20.6 (14.3) & 3.9 (4.4) & (9.4)  \\
&&&&\\
0.004/NoWind  & 50.0 (50.0) & 22.0 (18.9) & 4.2 (4.5) & (45.0) \\
0.004/Hurley  & 46.9 (46.5) & 21.6 (17.1) & 4.2 (4.6) & (6.7)  \\
0.004/Vink    & 47.4 (47.3) & 21.6 (17.5) & 4.2 (4.6) & (14.8) \\
&&&&\\
0.0003/NoWind & 50.0 (50.0) & 22.0 (18.9) & 4.3 (4.6) & (45.0) \\
0.0003/Hurley & 49.2 (49.1) & 22.0 (18.5) & 4.3 (4.6) & (8.6)  \\
0.0003/Vink   & 49.7 (49.6) & 22.0 (18.7) & 4.3 (4.6) & (17.9) \\
\enddata
\label{mod50}
\tablenotetext{a}{
Values for detailed evolutionary calculations (and the analytic formulae) are
given for a star with initial mass $M_{\rm zams}=50 \msun$. $M_{\rm tms}$--star mass 
at the end of Main sequence; $M_{\rm he}$--helium core mass at the end of MS;  
$\tau_{\rm ms}$--MS lifetime; $M_{\rm bh}$--final mass of a remnant (this is
given only for our calculations as evolutionary models were not evolved
beyond main sequence. For details see Sec.4.2. 
} 
\end{deluxetable}

\begin{deluxetable}{lcccc}
\tablewidth{420pt}
\tablecaption{Comparison of Models: $M_{\rm zams}=100 \msun$ \tablenotemark{a}}
\tablehead{
$Z$/Wind& $M_{\rm tms}$[$\msun$] & $M_{\rm he}$[$\msun$] & $\tau_{\rm ms}$[Myr] & $M_{\rm bh}$[$\msun$] }

\startdata
0.02/NoWind   & 100.0 (100)  & 51.9 (48.3) & 2.7 (3.4) & (90.0) \\
0.02/Hurley   & 63.0 (60.3) & 43.9 (24.4) & 2.7 (3.5) & (9.8)  \\
0.02/Vink     & 60.9 (59.1) & 43.1 (23.7) & 2.8 (3.5) & (13.7) \\
&&&&\\
0.004/NoWind  & 100.0 (100)  & 50.6 (48.3) & 2.9 (3.5) & (90.0) \\
0.004/Hurley  & 86.2 (80.5) & 47.3 (36.0) & 2.9 (3.5) & (11.6) \\
0.004/Vink    & 94.2 (84.0) & 49.3 (38.1) & 2.9 (3.5) & (33.5) \\
&&&&\\
0.0003/NoWind & 100.0 (100)  & 50.6 (43.4) & 3.0 (3.5) & (90.0) \\
0.0003/Hurley & 96.5 (95.1) & 49.4 (41.0) & 3.0 (3.5) & (12.4) \\
0.0003/Vink   & 99.5 (98.2) & 50.5 (42.5) & 3.0 (3.5) & (42.9) \\
\enddata
\label{mod100}
\tablenotetext{a}{
Same as Table~\ref{mod50} but for initial star with mass $M_{\rm zams}=100 \msun$.
} 
\end{deluxetable}

\clearpage

\clearpage

\begin{figure}
\includegraphics[width=1.0\columnwidth]{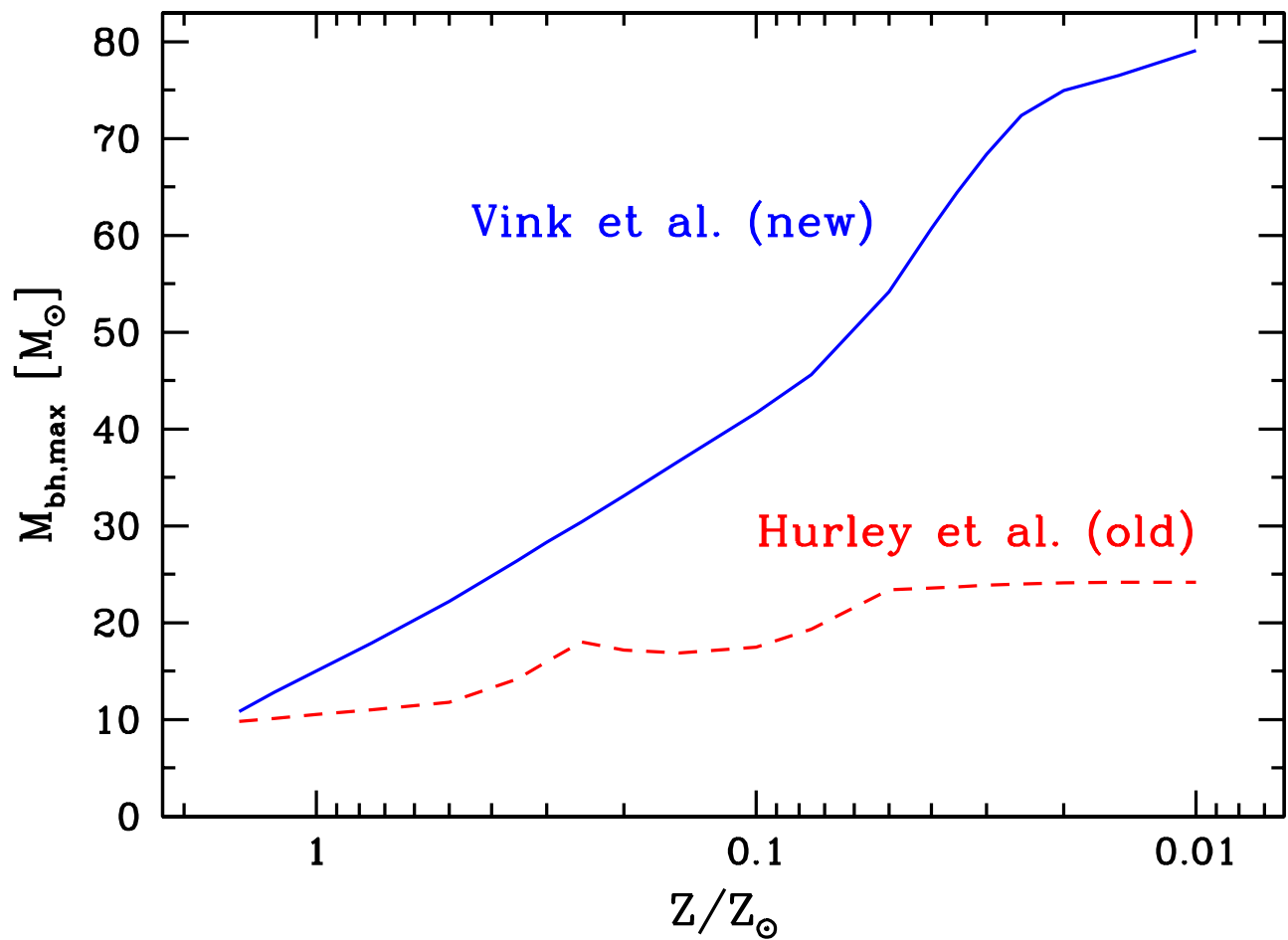}
\caption{
The dependence of maximum black hole mass on metallicity for our previously
used set of stellar winds (Hurley et al.; \S\,2.1) and for new adopted winds 
(Vink et al.; \S\,2.2). Note that the maximum BH mass is similar for solar-like
metallicities, while it is significantly larger for the new winds in metal-poor
environments.
}
\label{bhmax}
\end{figure}

\end{document}